\documentclass[pra,twocolumn,amsmath,amssymb,superscriptaddress]{revtex4}
\usepackage[paperwidth=210mm,paperheight=297mm,centering,hmargin=2.0cm,vmargin=2.6cm]{geometry}
\usepackage{graphicx}
\usepackage{dcolumn}
\usepackage{bm}
\usepackage{upgreek}
\usepackage{amsmath}
\usepackage{url}
\usepackage{hyperref}
\usepackage{lipsum}

\newcommand{\beq}{\begin{equation}}
\newcommand{\eeq}{\end{equation}}
\newcommand{\bea}{\begin{align}}
\newcommand{\eea}{\end{align}}
\newcommand{\beqa}{\begin{eqnarray}}
\newcommand{\eeqa}{\end{eqnarray}}
\newcommand{\e}{\mathrm{e}}
\newcommand{\w}{\omega}
\newcommand{\ket}[1]{\left| #1 \right\rangle}

\newcommand{\av}[1]{\langle #1\rangle}

\newcommand{\der}[2]{\frac{\mathrm{d}#1}{\mathrm{d}#2}}
\newcommand{\ketbra}[2]{\left|#1\right\rangle\hskip-1mm\left\langle #2\right|}

\newcommand{\sfrac}[2]{{\textstyle{\frac{ #1}{#2}}}}
\newcommand{\gtot}{\gamma_{\mathrm{tot}}}
\newcommand{\pder}[2]{\frac{\partial #1}{\partial #2} }

\begin{document}
\title{Two-photon interference from a quantum dot--microcavity: 
Persistent pure-dephasing and suppression of time-jitter}

\author{Sebastian Unsleber}
\affiliation{Technische Physik and Wilhelm Conrad R\"ontgen Research Center for Complex Material Systems, Physikalisches Institut,
Universit\"at W\"urzburg, Am Hubland, D-97074 W\"urzburg, Germany}
\author{Dara P. S. McCutcheon}
\affiliation{Department of Photonics Engineering, Technical University of Denmark, \O rsteds Plads, 2800 Kgs. Lyngby}
\author{Michael Dambach}
\affiliation{Technische Physik and Wilhelm Conrad R\"ontgen Research Center for Complex Material Systems, Physikalisches Institut,
Universit\"at W\"urzburg, Am Hubland, D-97074 W\"urzburg, Germany}
\author{Matthias Lermer}
\affiliation{Technische Physik and Wilhelm Conrad R\"ontgen Research Center for Complex Material Systems, Physikalisches Institut,
Universit\"at W\"urzburg, Am Hubland, D-97074 W\"urzburg, Germany}
\author{Niels Gregersen}
\affiliation{Department of Photonics Engineering, Technical University of Denmark, \O rsteds Plads, 2800 Kgs. Lyngby}
\author{Sven H\"ofling}
\affiliation{Technische Physik and Wilhelm Conrad R\"ontgen Research Center for Complex Material Systems, Physikalisches Institut,
Universit\"at W\"urzburg, Am Hubland, D-97074 W\"urzburg, Germany}
\affiliation{SUPA, School of Physics and Astronomy, University of St Andrews, St Andrews, KY16 9SS, United Kingdom}
\author{Jesper M\o rk}
\affiliation{Department of Photonics Engineering, Technical University of Denmark, \O rsteds Plads, 2800 Kgs. Lyngby}
\author{Christian Schneider}
\affiliation{Technische Physik and Wilhelm Conrad R\"ontgen Research Center for Complex Material Systems, Physikalisches Institut,
Universit\"at W\"urzburg, Am Hubland, D-97074 W\"urzburg, Germany}
\author{Martin Kamp}
\affiliation{Technische Physik and Wilhelm Conrad R\"ontgen Research Center for Complex Material Systems, Physikalisches Institut,
Universit\"at W\"urzburg, Am Hubland, D-97074 W\"urzburg, Germany}

\date{\today}

\begin{abstract}

We demonstrate the emission of highly indistinguishable photons from a quasi-resonantly pumped coupled quantum dot--microcavity system operating 
in the regime of cavity quantum electrodynamics. Changing the sample temperature allows us to vary the quantum dot--cavity detuning, and 
on spectral resonance we observe a three-fold improvement in the Hong--Ou--Mandel interference visibility, 
reaching values in excess of 80\%. Our measurements off-resonance allow us to investigate varying Purcell enhancements, and to probe 
the dephasing environment at different temperatures and energy scales. By comparison with our microscopic model, we are able 
to identify pure-dephasing and not time-jitter as the dominating source of imperfections in our system.

\end{abstract}
\maketitle

\section{Introduction}
Single indistinguishable photons are key to applications in quantum networks~\cite{Pan2012}, linear 
optical quantum computing~\cite{Kok2007,OBrien2007} and quantum teleportation~\cite{Nilsson-NatPhot13, Gao-NatCom13}. 
One of the most promising platforms for single photon sources are solid-state quantum dots 
(QDs)~\cite{Michler2000,Santori2002,Flagg2010,Gold2014,Muller2014}. 
Compared to alternative platforms, such as cold atoms or trapped ions, single 
QDs offer several advantages: they can be driven electrically, which is of crucial importance for compact future 
applications~\cite{Yuan2002, Heindel2010, Ellis2008}, and in principle can be integrated in complex photonic environments 
and architectures, such as on-chip quantum optical networks~\cite{yao09,Hoang12}. When embedded in a 
bulk semiconductor, however, QDs suffer from poor photon extraction efficiencies, since only a minor 
fraction of the photons can leave the high refractive index material. 
This problem can be mitigated by integrating QDs into optical microcavities~\cite{Heindel2010,Reitzenstein2010,gazzano13,Gerard1998} 
or photonic waveguides~\cite{Claudon2010a,Reimer2012,arcari14}, which can enhance extraction efficiencies to values beyond 50\%. 

In addition to increased extraction efficiencies, exploiting cavity quantum electrodynamics (cQED) effects in 
QD-based sources can have a positive effect on the interference properties 
(and hence the indistinguishability) of the emitted photon wave packets. 
Ideally, the wave packets emitted by an indistinguishable photon source are Fourier-limited, 
with a recombination time $T_1$, and temporal extension 
of the wave packet given by $T_2=2 T_1$ \cite{He2013}. If additional dephasing channels with a characteristic time $1/\gamma$ exist, such as 
coupling to phonons or spectral diffusion, the coherence time is reduced according to 
$\smash{\frac{1}{T_2}=\frac{1}{2 T_1}+\gamma}$, which consequently leads to a reduction of the two photon interference visibility. 
It was theoretically shown that pure dephasing strongly affects the detuning dependence of the relative strength of the cavity and QD-emission 
peaks~\cite{Naesby-PRA08, Auffeves-PRA09}.
In the regime of cQED, 
the lifetime of the QD excitons can be manipulated via the photonic density of states in the cavity (the Purcell effect). 
If the timing of emission events is precisely known, and $\gamma$ is constant, shortening of the emitter 
lifetime $T_1$ via the Purcell effect leads to an improved interference visibility as the condition $T_2=2 T_1$ can be 
approximately restored~\cite{Santori2002,gazzano13,Varoutsis2005}. 
This simple picture, however, is known to breakdown if there are uncertainties in the timing of 
emission events (time-jitters)~\cite{kiraz04,Troiani2006,Kaer2013}, or if the dephasing environment gives rise to more 
than a simple constant pure-dephasing rate, as is known to be the case for phonons~\cite{Kaer2013,Kaer13_2,Kaer2014,mccutcheon13,Wei2014}. 
As such, with the aim of designing improved single indistinguishable photon sources, 
it is crucially important to first establish the magnitude of time-jitters and the 
nature of any dephasing environments.

In this work, we exploit a microcavity with a high Purcell factor and weak non-resonant contributions of spectator QDs to 
probe the interference properties of photons emitted from a single QD as a function of the QD--cavity detuning. 
In contrast to previous studies, where non-resonant coupling to spectator QDs~\cite{Weiler-PSSB08} or strong temperature induced 
dephasing~\cite{Varoutsis2005} dominated the experiments, we observe a strong improvement of the two-photon visibility on 
resonance, which exceeds a factor of 3 compared to the off-resonant case. We extend the theoretical model of Ref.~\cite{Kaer2013} to 
derive an expression for the Hong--Ou--Mandel dip including the effects of both time-jitter and pure-dephasing on- and off-resonance. 
This allows us to reject timing-jitter, 
and definitively attribute sources of pure-dephasing as the dominant factor limiting the indistinguishability of our photons. Furthermore, 
we show that the degree of symmetry we observe for positive and negative detuning suggests pure-dephasing caused by both 
phonon coupling and spectral diffusion.

\section{Quantum dot--cavity system}

The device under investigation comprises a QD embedded in a micropillar cavity with a  quality factor of $Q=3200$. 
The layer structure consists of 25 (30) alternating 
$\frac{\lambda_C}{4*n}$-thick GaAs/AlAs mirror pairs which form the upper (lower) distributed Bragg reflector (DBR). 
The cavity region is composed of six alternating GaAs/AlAs layers with decreasing (lower part) and increasing (upper part) thickness. 
A single layer of partially capped and annealed InAs QDs is integrated in the central layer of the tapered segment, i.e. in the vertical 
maximum of the optical field~\cite{Lermer2012}. 
Micropillars with varying diameters were etched into the wafer 
(the pillar under investigation has a diameter of $1050~\mathrm{nm}$) to provide zero dimensional mode confinement. 
As a result of the Bloch mode engineering~\cite{Lermer2012}, our micropillars support optical resonances with comparably large quality 
factors down to the sub-micron diameter range, which yields the possibility to significantly increase the Purcell factor in such 
microcavities compared to conventional DBR resonators based on $\lambda$-thick cavity spacers. 
The sample was placed inside an optical cryostat, and the QD was excited via a picosecond-pulsed Ti:sapphire laser 
with a repetition frequency of $82$ MHz (pulse separation $12.2~\mathrm{ns}$). 
The laser beam was coupled into the optical path via a polarizing beam splitter, 
which also suppresses the scattered laser light from the detection path of the setup. Further filtering was implemented by a long-pass filter in 
front of the monochromator. After spectral filtering, the emitted photons were coupled into a polarization maintaining single mode 
fibre followed by a fibre coupled Mach-Zehnder-Interferometer (MZI) with a variable fibre-coupled time delay in one arm 
to measure the two-photon-interference in a 
Hong--Ou--Mandel (HOM) setup. The second beamsplitter of the MZI can be removed to directly measure the autocorrelation function of the signal.

\begin{figure}
\begin{center}
\includegraphics[width=0.48\textwidth]{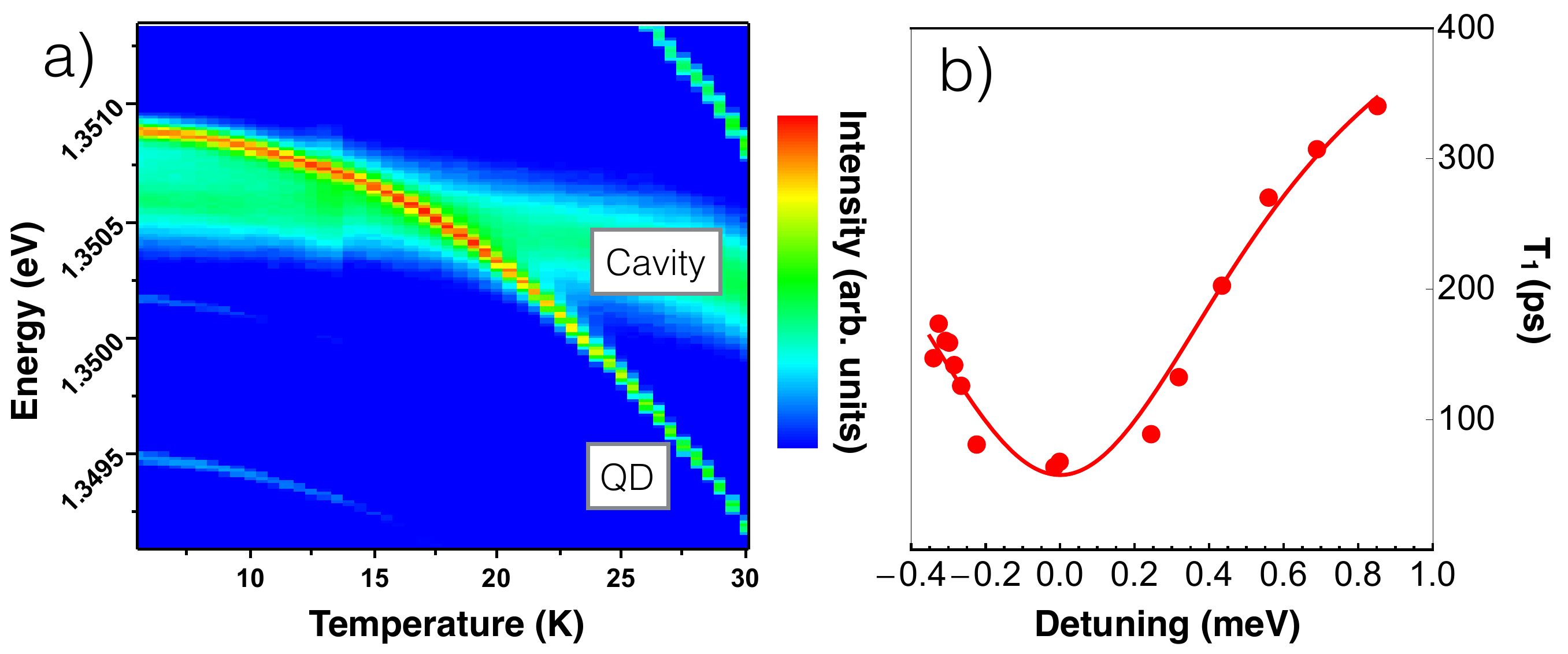}
\caption{a) Temperature dependent intensity map of the QD--cavity system. 
The QD reaches spectral resonance with the cavity mode at $T=17.5$ K. b) QD lifetime as a function of QD--cavity detuning. 
The fit is a Lorentzian profile where the linewidth is fixed to the cavity linewidth. 
A Purcell enhancement of $F_P=7.8\pm2.3$ is extracted.}
\label{TempMapAndT1}
\end{center}
\end{figure}

Fig.~{\ref{TempMapAndT1}} (a) shows the temperature dependent micro-photoluminescence ($\mu$-PL) map of the investigated 
QD--cavity system, which was recorded under non-resonant excitation conditions. The QD emission line, which we attribute to the neutral exciton, 
can be tuned through the cavity mode by changing the sample temperature. 
Spectral resonance with the fundamental cavity mode is achieved at $T = 17.5$~K. Due to the Purcell enhancement, 
the integrated intensity of the QD increases by a factor of more than three when the QD and cavity are tuned into resonance. 
In order to directly and accurately extract the Purcell factor of our coupled system, we measured the exciton lifetime via time-resolved 
$\mu$-PL as a function of the QD-cavity detuning (see Appendix~\ref{lifetime_measurement}). 
As seen in Fig.~\ref{TempMapAndT1} (b), we observe a strong decrease of the lifetime when the QD 
is tuned into resonance as a result of the Purcell effect. The Purcell factor 
$F_P=\smash{\frac{T_1(\Delta\to\infty)}{T_1(\Delta\to 0)}-1}$~\cite{Munch-PRB} is extracted by fitting 
the data with a Lorentzian profile (the width being fixed to the cavity linewidth $\kappa=0.42~\mathrm{meV}$), and we find a value as 
high as $F_P=7.8\pm2.3$ as a result of the small mode volume of our microcavity.

\begin{figure}[b]
\begin{center}
\includegraphics[width=0.48\textwidth]{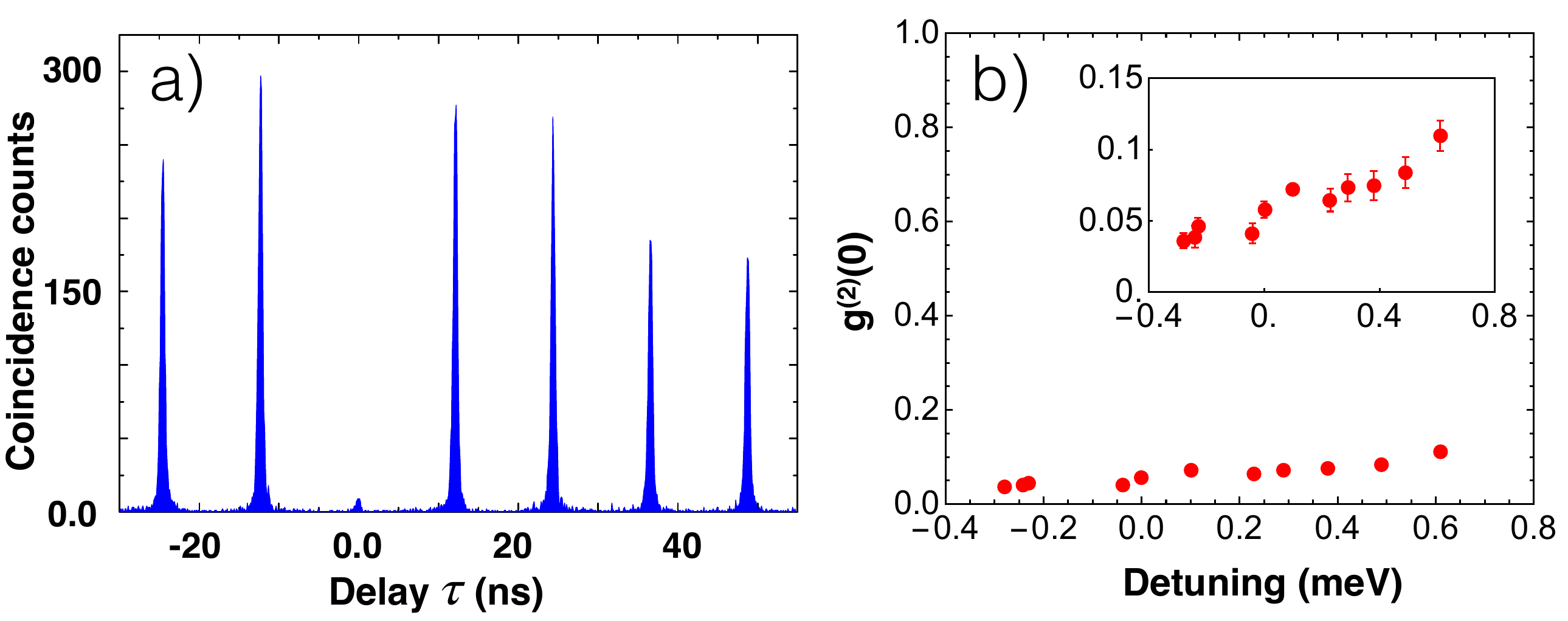}
\caption{a) Autocorrelation histogram on QD--cavity resonance, from which we 
extract $\smash{g^{(2)}(0) = (0.058 \pm 0.006)}$. 
b) Autocorrelation function as a function of QD--cavity detuning (the inset shows a zoom-in of the y-axis).}
\label{Autocorrelation}
\end{center}
\end{figure}

We now study the single photon emission properties of our system, 
which is particularly important on-resonance, where the single photon 
characteristics can be deteriorated by non-resonant contributions to the cavity from spectator 
QDs, or luminescence from the background continuum funnelled into the cavity mode~\cite{Hennessy-Nature07}. 
The second-order photon-autocorrelation was probed under quasi-resonant excitation conditions, 
with a laser tuned $32~\mathrm{meV}$ to the high energy side of the single exciton emission feature, 
with a (below saturation) power of $311~\mu\mathrm{W}$. The on-resonance ($T = 17.5$~K) autocorrelation 
histogram is shown in Fig.~{\ref{Autocorrelation}} (a). 
The strongly suppressed peak around $\tau = 0$ is a clear signature of single photon emission. 
We extract the $g^{(2)}(\tau=0)$ value by dividing the area of the central peak by the average area of all the side peaks, leading to 
$g^{(2)}(0) = 0.058 \pm 0.006$, reflecting the high purity of our cavity enhanced single photon source. 
Off-resonance we find a minimum value of 
$\smash{g^{(2)}(0)}=(0.036 \pm 0.005)$ at 
$\Delta = -0.28$ meV ($T = 6.4$~K). For increasing temperatures, we note a modest increase up to 
$g^{(2)}(0)  = (0.11 \pm 0.01)$ for $\Delta = 0.61$ meV ($T = 25.5$~K). 
This value is still close to perfect single photon emission, 
and we attribute the slight rise to a lowered signal to background ratio between QD and cavity emission. We note that 
no deterioration of the $g^{(2)}(0)$ value can be observed on spectral resonance, which suggests only very weak contributions from 
spectator QDs to the cavity signal. 

\begin{figure}
\begin{center}
\includegraphics[width=0.48\textwidth]{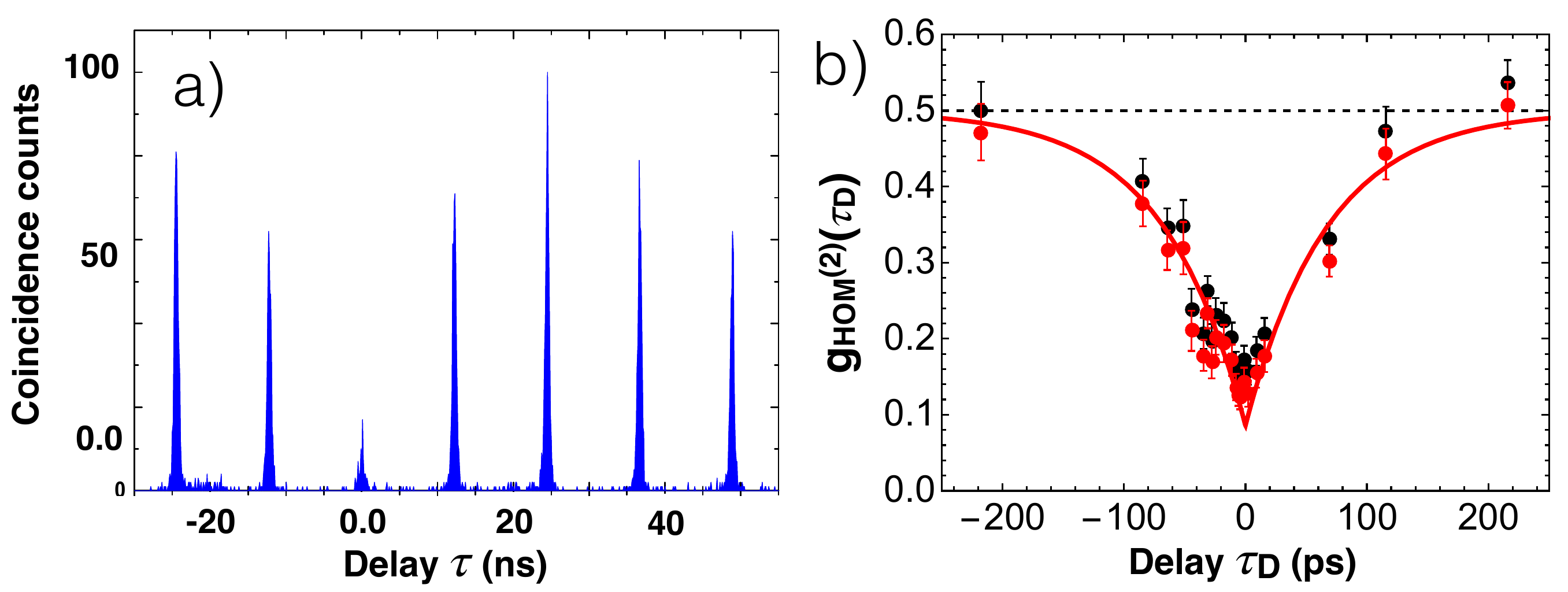}
\caption{a) Histogram of the two-photon-interference for zero time delay between the two arms of the interferometer, from which 
we extract $\smash{g^{(2)}_{\mathrm{HOM}}}(0)= 0.17 \pm0.02$. b) Measured two-photon-interference versus the time delay. 
The measurement shows the clear HOM-dip (black dots). Due to a non-vanishing two-photon probability, the data points go slightly 
above $0.5$ for large $\tau_D$. These are corrected by subtracting half the measured $\smash{g^{(2)}(0)}$ value on resonance 
shown in Fig.~\ref{Autocorrelation} 
(red markers). These corrected data were fitted to extract a visibility of $\nu=(83\pm5)$\%.}
\label{HOM}
\end{center}
\end{figure}

\section{Photon Indistinguishability}

We now assess the 
indistinguishable nature of the emitted photons, which we probe in the HOM interferometer under the same pulsed quasi-resonant excitation conditions. 
The second order correlation 
histogram for zero time delay in the MZI for the resonant case is shown in Fig.~\ref{HOM} (a). Strong suppression of the central correlation 
peak directly reflects a strong degree of photon indistinguishability. The black markers in Fig.~{\ref{HOM}}~(b) are obtained 
by dividing the area of the peak centred around $\tau=0$ by that centred around $\tau=24~\mathrm{ns}$ for various time 
delays $\tau_D$, and we observe a clear HOM-dip. 
For large time delays $\tau_D$ the correlation values slightly exceed $0.5$ as a result of the finite two photon emission 
probability, as seen in Fig.~\ref{Autocorrelation}. We correct the interference data by subtracting half the corresponding experimentally 
extracted on-resonance value of $g^{(2)}(0)=(0.058 \pm 0.006)$ [red markers] (see Appendix~\ref{HOMExperiment}). We 
then fit our data to the function 
$\smash{g_{\mathrm{HOM}}^{(2)} (\tau_D)=0.5(1- \nu \exp[-|\tau_D|/T_1])}$, 
where we set $T_1=67~\mathrm{ps}$ (see Fig.~\ref{TempMapAndT1} (b)), and we find a visibility of $\nu=83\pm5\%$. This high value is a direct 
consequence of the large Purcell factor in our high quality QD--cavity system.

To further analyse our experimental data, and in particular, to determine the relative 
influences of time-jitter and pure-dephasing on the indistinguishability of the emitted photons, we 
extend the theory of Ref.~[\onlinecite{Kaer2013}] to derive an expression for the TPI as a function of 
both time delay $\tau_D$ and detuning. Dephasing caused by coupling to phonons is known 
to affect the two-photon interference (TPI) properties of the emission from a QD--cavity system in a highly non-trivial way, giving 
rise, for example, to pronounced asymmetries for positive and negative QD--cavity 
detunings~\cite{Kaer2013,Kaer13_2,Kaer2014}. We find, however, that nearly all 
features seen in our data can be well reproduced by a model assuming a simple constant pure-dephasing rate. 
We present this simplified model first, and then go on to show that by including phonons in a rigorous manner at 
a Hamiltonian level, the behaviour off-resonance allows us to approximately determine the relative influence of 
phonons as compared to other sources of dephasing.

We model the QD as a three-level-system, and consider the 
vacuum and single photon Fock states of the cavity. Provided the QD--cavity coupling strength is sufficiently weak, 
and/or the cavity decay rate is sufficiently large, the cavity degrees of freedom can be adiabatically eliminated 
from the equations of motion for the QD--cavity system~\cite{Kaer2013}. The result is a master equation of the form (see 
Appendix~\ref{QDCS})
\begin{align}
&\der{\rho}{t}=-\sfrac{i}{\hbar}[\Delta\ketbra{E}{E},\rho]+ \nonumber \\
&\big(L_{\Gamma}(\ketbra{G}{E})
+L_{2\gamma}(\ketbra{E}{E})+L_{\alpha}(\ketbra{E}{P})\big)\rho,
\label{MEFinal}
\end{align}
where the states $\ket{E}=\ket{\e,n=0}$, $\ket{G}=\ket{\mathrm{g},n=1}$ $\ket{P}=\ket{\mathrm{p},n=0}$ represent 
the QD in ground (g), single exciton state (e), or pump-level (p), with the cavity containing 
zero or one excitations. The QD--cavity detuning is $\Delta$, while $\gamma$ is the pure-dephasing rate, 
and $\alpha$ is the rate at which the pump-level decays into the single exciton state, with $T_{\alpha}=1/\alpha$
determining the magnitude of the time-jitter (i.e. $T_{\alpha}=0$ represents the ideal case in which there 
is no time-jitter). The Purcell enhanced spontaneous emission rate is 
\beq
\Gamma=T_1^{-1}=\Gamma_B+2 g^2\frac{\gtot}{\gtot^2+\Delta^2},
\label{PurcellR}
\eeq
with $\Gamma_B$ the background decay rate, $g$ the QD--cavity coupling strength, 
and $\gtot=\gamma+\frac{1}{2}(\kappa+\Gamma_B)$ with 
$\kappa$ the cavity decay rate. The validity of Eq.~({\ref{MEFinal}}) relies on the condition 
$\gtot\gg\Delta,\Gamma,g$, which is satisfied in all our experiments.

Eq.~({\ref{MEFinal}}) can then be used to derive an expression for the normalised coincidence 
events in the TPI measurements (for details see Appendix~\ref{QDCS}). 
The second order correlation function 
for the HOM interference measurements is found to read 
\begin{equation}
g^{(2)}_{\mathrm{HOM}}(\tau_D)=\frac{1}{2}\Big(1-\frac{\nu}{\Gamma-\alpha}\Big[\Gamma\e^{-|\tau_D|\alpha}-\alpha\e^{-|\tau_D|\Gamma}\Big]\Big)
\label{HOMDipalpha}
\end{equation}
where the detuning dependence enters through $\Gamma$ [see Eq.~({\ref{PurcellR}})], and 
$\nu=(\Gamma/(\Gamma+2\gamma))(\alpha/(\Gamma+\alpha))$ is the visibility. We note 
that while the expression for $\nu$ has been derived before~\cite{Kaer2013}, 
to our knowledge Eq.~({\ref{HOMDipalpha}}) represents the first time 
the full behaviour of the HOM-dip for nonzero values of $\tau_D$ including 
time-jitter and pure-dephasing has been presented. 
This model provides us with simple analytical expressions with which we can fit the experimental TPI data. Crucially, it 
allows us to explore how a given set of parameters simultaneously affects the HOM-dip \emph{and} the 
TPI visibility as the QD and cavity are moved off resonance.

\begin{figure}
\begin{center}
\includegraphics[width=0.48\textwidth]{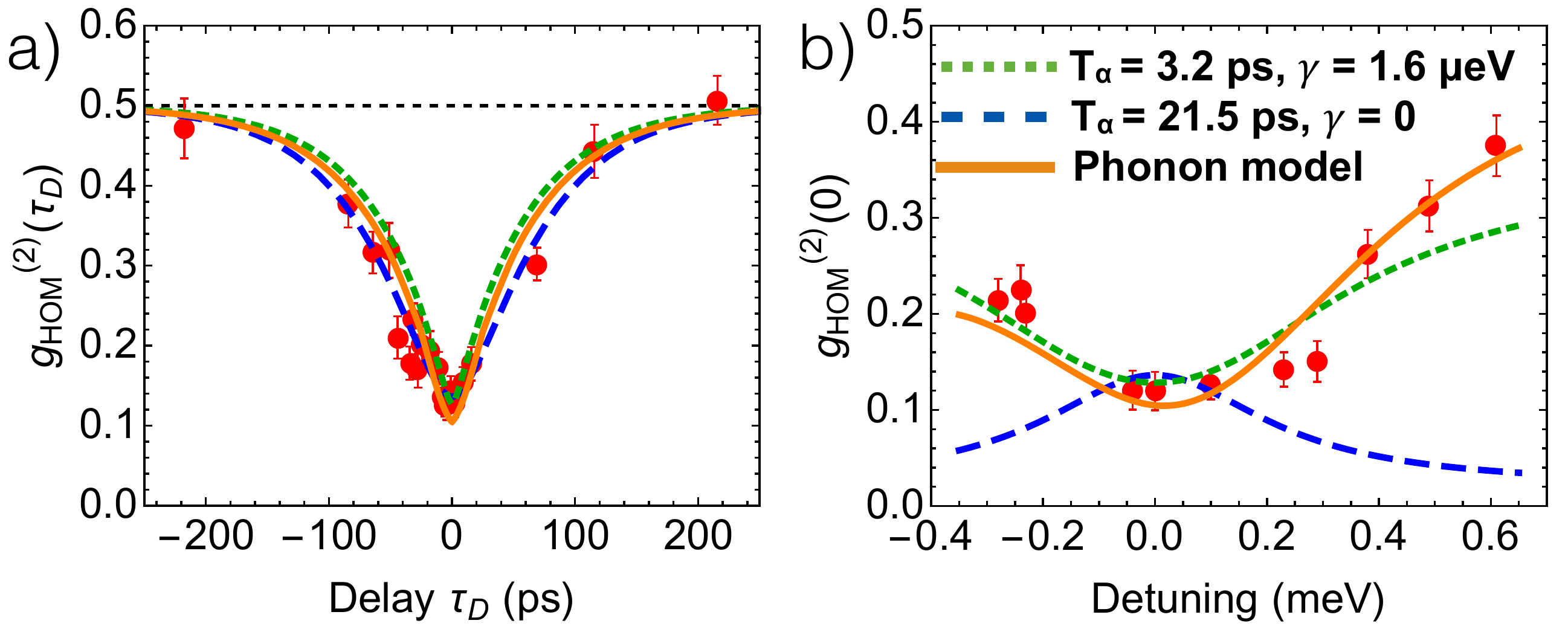}
\caption{a) HOM-dip and HOM-dip depth as a function of detuning (b). The dashed blue curves 
correspond to a parameter set for which time-jitter dominates, which we find to be inconsistent with the data 
off resonance shown in (b). The dotted green curves correspond to a parameters set for which pure-dephasing dominates, which 
is able to consistently reproduce all data. The solid orange curves again correspond to a parameter set dominated by pure-dephasing, but where 
$40\%$ of the dephasing on-resonance is caused by coupling to phonons.}
\label{HOMFits}
\end{center}
\end{figure}

In Fig.~{\ref{HOMFits}} (a) we show again the HOM-dip, while in Fig.~{\ref{HOMFits}} (b) 
we show the depth of HOM-dip as a function of detuning. We see a pronounced rise of the HOM-dip as the QD is 
brought out of resonance, corresponding to visibilities on- and off-resonance which differ by more than a factor of $3$. 
The dashed blue curves in Fig.~{\ref{HOMFits}} show a fit to Eq.~({\ref{HOMDipalpha}}) where 
only the data in Fig.~{\ref{HOMFits}} (a) (the HOM-dip) is considered. Fitting parameters $T_{\alpha}=1/\alpha=21.5~\mathrm{ps}$ and 
$\gamma=0~\mu\mathrm{eV}$ are found, corresponding to a regime where time-jitter dominates. These parameters are 
able to reproduce the HOM-dip well, but fail to describe the data off resonance. Indeed, we find that if we simultaneously fit 
all the data shown in Fig.~{\ref{HOMFits}}, we find parameters $T_{\alpha}=1/\alpha=3.2~\mathrm{ps}$ and $\gamma=1.6~\mu\mathrm{eV}$, 
corresponding to a regime for which pure-dephasing dominates. These parameters 
are shown by the dotted green curves, and much better agreement is found. 
These two fits show that while both time-jitter and pure-dephasing affect the shape of the HOM dip in a similar way, 
the reduction in the visibility seen off-resonance can only be explained by a system in which pure-dephasing dominates. 
As the QD and cavity are moved off resonance, the Purcell effect weakens (see Fig.~{\ref{TempMapAndT1}}) 
and $\Gamma=1/T_1$ decreases. For a source dominated by pure-dephasing, the visibility is given by 
$\nu\approx\Gamma/(\Gamma+2\gamma)$, and a reduction in $\Gamma$ causes a reduction in $\nu$. For a source dominated by 
time-jitter, the visibility instead follows $\nu\approx\alpha/(\Gamma+\alpha)$, and a reduction in $\Gamma$ increases $\nu$. 
We stress that which of two regimes is relevant for a particular system has important consequences for how experimental improvements 
will translate to improvements in photon indistinguishabilities. In the present case, since pure-dephasing dominates, 
a complete elimination of time-jitters (achieved, for example, via strictly resonant excitation conditions), will lead to only a modest 
$4\%$ increase in the visibility, while an elimination of sources of pure-dephasing will lead to an increase of $20\%$ up to $\nu=95\%$.

\section{Discussion}

The low value of $T_{\alpha}=3.2~\mathrm{ps}$ implies that our quasi-resonant excitation scheme leads to 
a very fast relaxation to the desired single exciton state. This is also supported by the laser detuning we 
use ($32~\mathrm{meV}$), which corresponds to the energy of a longitudinal optical phonon, known to relax on this timescale~\cite{Grange07}. 
We attribute pure-dephasing in our 
sample as caused by exciton--phonon coupling and spectral fluctuation of the QD energy levels on a timescale 
shorter than the pulse separation of $12.2~\mathrm{ns}$. 
The constant pure-dephasing rate used in our theory is expected to well approximate the spectral fluctuations, 
but the influence of phonons is known to give rise to more complicated behavior~\cite{Kaer2013,Kaer13_2,Kaer2014,mccutcheon13}. 
In particular, differing phonon absorption and emission rates at low temperatures are expected to lead to 
asymmetries for positive and negative detuning~\cite{Kaer13_2}. By including phonons 
using a time-convolutionless master equation technique~(see e.g. Ref.~\cite{Kaer2013} or Appendix~\ref{EPC}), 
we find that these asymmetries can improve our fits. 
The solid orange curves in Fig.~{\ref{HOMFits}} show the predictions of a parameter set similar to that of the dotted green curve, 
but where we have included phonons with a strength corresponding 
to approximately $40\%$ of the total pure-dephasing on-resonance~\footnote{We have also adjusted g and $\gamma_B$ so that the $T_1$ data are well reproduced  off resonance}, 
and it can be seen that the phonon contribution improves the fits to the data. 
We note, however, that when increasing the phonon contribution yet further, the fits become worse as the asymmetry becomes too strong. 
The relatively strong symmetry seen in Fig.~{\ref{HOMFits}} (b) therefore leads us to conclude 
that both phonons, and additional sources of constant pure-dephasing (such as a spectral diffusion) are present in our system.

In conclusion, we have demonstrated the feasibility of our novel cavity design to enhance the emission of indistinguishable 
single photons generated in epitaxially grown InAs-QDs by a quasi-resonant excitation scheme with a TPI-visibility as high as 
$\nu= (83\pm 5)$\%, and a two-photon emission probability as low as $g^{(2)}(0)  = (0.036 \pm 0.005)$. 
We studied the influence of the QD--cavity detuning on both the two-photon-probability and the degree of 
indistinguishability of the emitted photons. The TPI measurements are explained by our new theory which takes the 
QD-cavity-detuning, time-jitter and pure-dephasing into account, and which identifies sources of pure-dephasing as the ultimate 
factor limiting the indistinguishably of emitted photons.

\section*{Acknowledgements}

The authors would like to thank M. Emmerling and A. Wolf for sample preparation. 
We acknowledge financial support by the State of Bavaria and the German Ministry of 
Education and Research (BMBF) within the projects Q.com-H, the Chist-era project SSQN, as well as the Villum Fonden via the NATEC 
Centre of Excellence. 
This work was additionally funded by project SIQUTE (contract EXL02) of the European Metrology Research Programme (EMRP). 
The EMRP is jointly funded by the EMRP participating countries within EURAMET and the European Union. 
S.H. gratefully acknowledges support by the Royal Society and the Wolfson Foundation.

\appendix
\renewcommand\thefigure{\thesection.\arabic{figure}}
\setcounter{figure}{0}
\label{appendix}

\section{Experimental Methods}

\subsection{Coherence Measurements}
In addition to the Hong--Ou--Mandel (HOM) interference measurements, the coherence of the emitted photons was measured 
using a free-beam unbalanced Michelson interferometer. 
One mirror is mounted on a $300~\mathrm{mm}$ long linear stage which defines the path length difference between both 
optical arms, and using an additional implemented piezo crystal at one mirror, the contrast of the emitted 
photons is measured as a function of the path length difference. The measurements (black data points) are shown in 
Fig.~{\ref{MichelsonFig}}~(a) for a QD in spectral resonance with the cavity mode. Fitting these data points to a 
Gaussian function of the form $A+B*\exp[-(\pi/2)(\tau/T_2)^2]$ we extract a coherence 
time of $T_2=(93\pm3)~\mathrm{ps}$. This value is slightly lower than the coherence time extracted from the HOM-dip in 
Fig.~{\ref{HOM}}, for which $T_2 = 111~\mathrm{ps}$. 
We attribute this slight discrepancy to a long term spectral jitter which affects the QD emission 
energy on timescales which are longer than the pulse separation. In the HOM-measurement, only subsequently 
emitted photons separated by $12.2~\mathrm{ns}$ (the laser pulse separation) contribute to the measured 
indistinguishability, and hence the inferred coherence time of $T_2 = 111~\mathrm{ps}$. The HOM measurements therefore 
include an effective time filter. In contrast, the measurements made using the Michelson interferometer 
are time-integrated, and as such 
long-term drifts and spectral diffusion result in a deterioration of the extracted $T_2$ value~\cite{Santori2002, Kuhlmann2013, Gold2014}.

\subsection{Lifetime Measurements}
\label{lifetime_measurement}

\begin{figure}
\begin{center}
\includegraphics[width=0.48\textwidth]{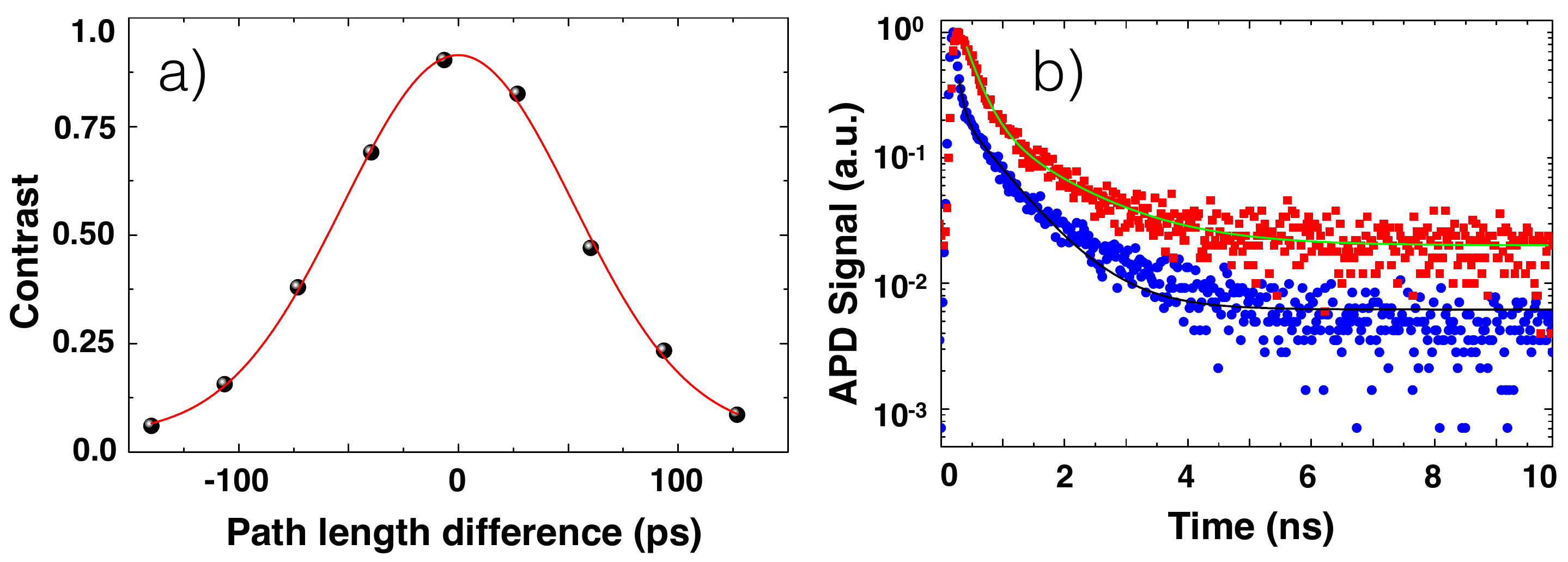}
\caption{a) Contrast of the emission measured using a free-beam unbalanced Michelson interferometer 
as a function of path length difference of both arms. Fitting the data points with a Gaussian distribution (red solid curve) we 
extract a coherence time of $T_2=(93\pm3)$ ps. 
b) Time-resolved $\mu$-PL measurements. 
The blue dotted data points correspond to spectral resonance between QD and fundamental cavity mode ($\Delta = 0$ meV), while 
the red square points were taken for a detuning of $\Delta=0.61$~meV. The solid curves 
correspond to fits to a biexponential function, from which we extract 
$T_1=(67\pm8)~\mathrm{ps}$ on resonance 
and $T_1=(306\pm13)~\mathrm{ps}$ off resonance.}
\label{MichelsonFig}
\end{center}
\end{figure}

In order to measure the lifetime of the QD emission we couple the spectrally filtered photons into a single 
mode fibre attached to an avalanche photo diode (APD) with resolution $\sim40~\mathrm{ps}$. 
Fig.~{\ref{MichelsonFig}}~(b) shows two representative time-resolved measurements of the QD emission under 
quasi-resonant excitation. The blue round 
data points correspond to spectral resonance between QD and fundamental cavity mode ($\Delta=0~\mathrm{meV}$), while 
the red square data points correspond to a detuning of $\Delta=0.61~\mathrm{meV}$. The measurements 
(time window $100$~ns) each contain six complete decay curves similar to those shown in Fig.~\ref{MichelsonFig}~(b), which we fit to a 
biexponential decay function. The shorter time constant represents the lifetime of the bright-exciton, while the longer  
originates from a dark exciton effect. For the decay curves in Fig.~{\ref{MichelsonFig}}~(b) we find $T_1=(67\pm8)~\mathrm{ps}$ on resonance 
and $T_1=(306\pm13)~\mathrm{ps}$ for $\Delta=0.61~\mathrm{meV}$. 

\section{Two-Photon Interference Theory}
\label{Theory}

Here we provide the necessary background for the theoretical 
analysis of the data presented in the main text.

\subsection{Hanbury Brown and Twiss Measurements}

We first consider the Hanbury Brown and Twiss (HBT) experimental setup used to measure 
the two-photon emission probability of our source. 
Emission from the source is incident upon 
a 50/50 beam splitter, and two detectors are placed equidistantly on the two output arms. We 
label $t_1$ the time of the detection event at detector 1, and 
$t_2$ that of detector 2. The probability of detecting a photon at detector 1 at $t_1$, and at detector 2 at $t_2$ is proportional 
to the second order field correlation function 
\beq
G^{(2)}(t_1,t_2)=\langle b_1^{\dagger}(t_1) b_2^{\dagger}(t_2) b_2(t_2) b_1(t_1)\rangle,
\label{G2Raw}
\eeq
where $b_1^{\dagger}(t)$ is the creation operator for the mode propagating to detector 1 in the Heisenberg picture, and similarly for 
$b_2^{\dagger}(t)$. We relate these modes to those on the input arms, described by creation operators $\smash{a_1^{\dagger}}$ and 
$\smash{a_2^{\dagger}}$, using the unitary mode transformation~\cite{kiraz04}
\beq
\left(
\begin{array}{c}
b_1^{\dagger}(t) \\ b_2^{\dagger}(t) \end{array}
\right)
=\frac{1}{\sqrt{2}}\left(\begin{array}{cc} 1 & 1 \\
-1 & 1 \end{array}\right)\left(\begin{array}{c} a_1^{\dagger}(t-\tau_D) \\ a_2^{\dagger}(t)\end{array}\right),
\label{BeamSplitter}
\eeq
where $\tau_D$ is the delay introduced between arrival times at the beam-splitter. 
For the HBT measurement, there is no input in arm 1, and we simply have $G^{(2)}(t_1,t_2)\to G_{\mathrm{HBT}}^{(2)}(t_1,t_2)$ with 
\beq
\begin{split}
G_{\mathrm{HBT}}^{(2)}(t_1,t_2)=
\frac{1}{4}\langle a^{\dagger}(t_1) a^{\dagger}(t_2) a(t_2) a(t_1)\rangle,
\end{split}
\label{G2HBT}
\eeq
where the subscripts on the operators have been dropped since they are all equal.

To measure the two-photon emission probability, $\smash{g^{(2)}(0)}$, we integrate Eq.~({\ref{G2HBT}}) over 
all $t_1$ and $t_2$, and divide this area by an adjacent peak. The adjacent peaks correspond to Eq.~({\ref{G2HBT}}), but 
where $t_1$ and $t_2$ differ sufficiently that mode operators at these times are completely uncorrelated. This 
gives the uncorrelated coincidence probability in the HBT measurement 
$\smash{G_{\mathrm{HBT,UC}}^{(2)}(t_1,t_2)=(1/4)\mathcal{G}^{(2)}(t_1,t_2)}$ with 
\beq
\mathcal{G}^{(2)}(t_1,t_2)=\langle a^{\dagger}(t_1) a(t_1)\rangle \langle a^{\dagger}(t_2) a(t_2)\rangle.
\eeq
The normalised autocorrelation function is then defined as 
\begin{align}
&g^{(2)}(0)=\frac{\int_{-\infty}^{\infty} d t_1 \int_{-\infty}^{\infty}d t_2 G_{\mathrm{HBT}}^{(2)}(t_1,t_2)}
{\int_{-\infty}^{\infty} d t_1 \int_{-\infty}^{\infty}d t_2 G_{\mathrm{HBT,UN}}^{(2)}(t_1,t_2)} \nonumber \\
&=\frac{\int_{-\infty}^{\infty} d t_1 \int_{-\infty}^{\infty}d t_2 \langle a^{\dagger}(t_1) a^{\dagger}(t_2) a(t_2) a(t_1)\rangle}
{\int_{-\infty}^{\infty} d t_1 \int_{-\infty}^{\infty}d t_2 \langle a^{\dagger}(t_1) a(t_1)\rangle \langle a^{\dagger}(t_2) a(t_2)\rangle}
\end{align}
which is equal to zero for $\langle a^{\dagger}(t_1) a^{\dagger}(t_2) a(t_2) a(t_1)\rangle=0$.

\subsection{Hong-Ou-Mandel Experiment}
\label{HOMExperiment}

We now consider the Hong--Ou--Mandel (HOM) experimental setup used to measure the indistinguishable nature of the 
emitted photons. Two emission 
events are incident on a 50/50 beam-splitter, with a delay $\tau_D$ introduced into input arm one. 
The unnormalised probability of a coincidence event is again given by Eq.~({\ref{G2Raw}}), and the beam-splitter is 
described by Eq.~({\ref{BeamSplitter}}). 
Upon combining these equations we find 16 terms. 
These can be simplified by assuming that modes 1 and 2 are identical but statistically independent, which 
allows us to write $\langle A_1 A_2\rangle=\av{A_1}\av{A_2}$, where $A_1$ is any product of mode operators pertaining to mode 1, and 
similarly for $A_2$. We then find eight terms linear in $\av{a_1}$ and $\av{a_2}$. 
For an electromagnetic field state of the form $\sum_n a_n \ketbra{n}{n}$, with $\ket{n}$ a Fock state, expectation values 
linear in the ladder operators are zero, and we neglect these terms. This leaves second and fourth order terms. 
The second order terms involve expectation values of the form $\langle a^{\dagger}(t_1) a^{\dagger}(t_2)\rangle$, which 
also give zero for electromagnetic fields as discussed above. The remaining terms give  
\begin{widetext}
\begin{align}
G^{(2)}_{\mathrm{HOM}}(t_1,t_2,\tau_D)=&\,G^{(2)}_{\mathrm{HBT}}(t_1-\tau_D,t_2-\tau_D)+G^{(2)}_{\mathrm{HBT}}(t_1,t_2) \nonumber\\
+&\frac{1}{4}\bigg(\mathcal{G}^{(2)}(t_1-\tau_D,t_2)+
\mathcal{G}^{(2)}(t_1,t_2-\tau_D)-
2\mathrm{Re}\left[G^{(1)}(t_1-\tau_D,t_2-\tau_D)G^{(1)}(t_2,t_1)\right]\bigg)
\label{G2HOM2}
\end{align}
\end{widetext}
where $G^{(1)}(t_1,t_2)=\av{a^{\dagger}(t_1) a(t_2)}$ is the unnormalised first order correlation function.

To normalise this quantity we again consider the scenario in which $t_1$ and $t_2$ are sufficiently separated that 
mode operators evaluated at these two times are uncorrelated. In doing so we find the uncorrelated coincidence probability 
for the HOM setup 
$G^{(2)}_{\mathrm{HOM,UC}}(t_1,t_2)=
\frac{1}{4}(\mathcal{G}^{(2)}(t_1,t_2)+\mathcal{G}^{(2)}(t_1-\tau_D,t_2)+\mathcal{G}^{(2)}(t_1,t_2-\tau_D)
+\mathcal{G}^{(2)}(t_1-\tau_D,t_2-\tau_D))$. 
Since we integrate over all $t_1$ and $t_2$, the appearances of $\tau_D$ 
can be neglected, i.e. we have 
\begin{align}
\int_{-\infty}^{\infty}d t_1\int_{-\infty}^{\infty}d t_2 &G^{(2)}_{\mathrm{HOM,UC}}(t_1,t_2)=\nonumber\\
&\int_{-\infty}^{\infty}d t_1\int_{-\infty}^{\infty}d t_2 \mathcal{G}^{(2)}(t_1,t_2)
\end{align}
by a simple change of variables. An identical argument can be made for the $\tau_D$ appearing in 
$G^{(2)}_{\mathrm{HBT}}(t_1-\tau_D,t_2-\tau_D)$ in Eq.~({\ref{G2HOM2}}). 
In the HOM setup we therefore measure the normalised quantity
\begin{align}
g^{(2)}_{\mathrm{HOM}}(\tau_D)&=\frac{\int_{-\infty}^{\infty} d t_1 \int_{-\infty}^{\infty}d t_2 G_{\mathrm{HOM}}^{(2)}(t_1,t_2,\tau_D)}
{\int_{-\infty}^{\infty} d t_1 \int_{-\infty}^{\infty}d t_2 G_{\mathrm{HOM,UN}}^{(2)}(t_1,t_2)}\nonumber\\
&=
\frac{1}{2}g^{(2)}(0)+\frac{1}{2}\Big(1-C(\tau_D)\Big)
\label{g2HOM}
\end{align}
where we have defined the strictly-two-photon coalescence probability 
\begin{align}
&C(\tau_D)=\nonumber\\
&\frac{\int_{-\infty}^{\infty}\! d t_1\! \int_{-\infty}^{\infty}\!d t_2 \mathrm{Re}\big[G^{(1)}(t_1-\tau_D,t_2-\tau_D)G^{(1)}(t_2,t_1)\big]}
{\int_{-\infty}^{\infty} d t_1 \int_{-\infty}^{\infty}d t_2 \mathcal{G}^{(2)}(t_1,t_2)}
\label{Ctau}
\end{align}
which for $\tau_D=0$ becomes the visibility $\nu=C(0)$.

\subsection{Quantum dot--cavity system}
\label{QDCS}

\begin{figure}
\begin{center}
\includegraphics[width=0.48\textwidth]{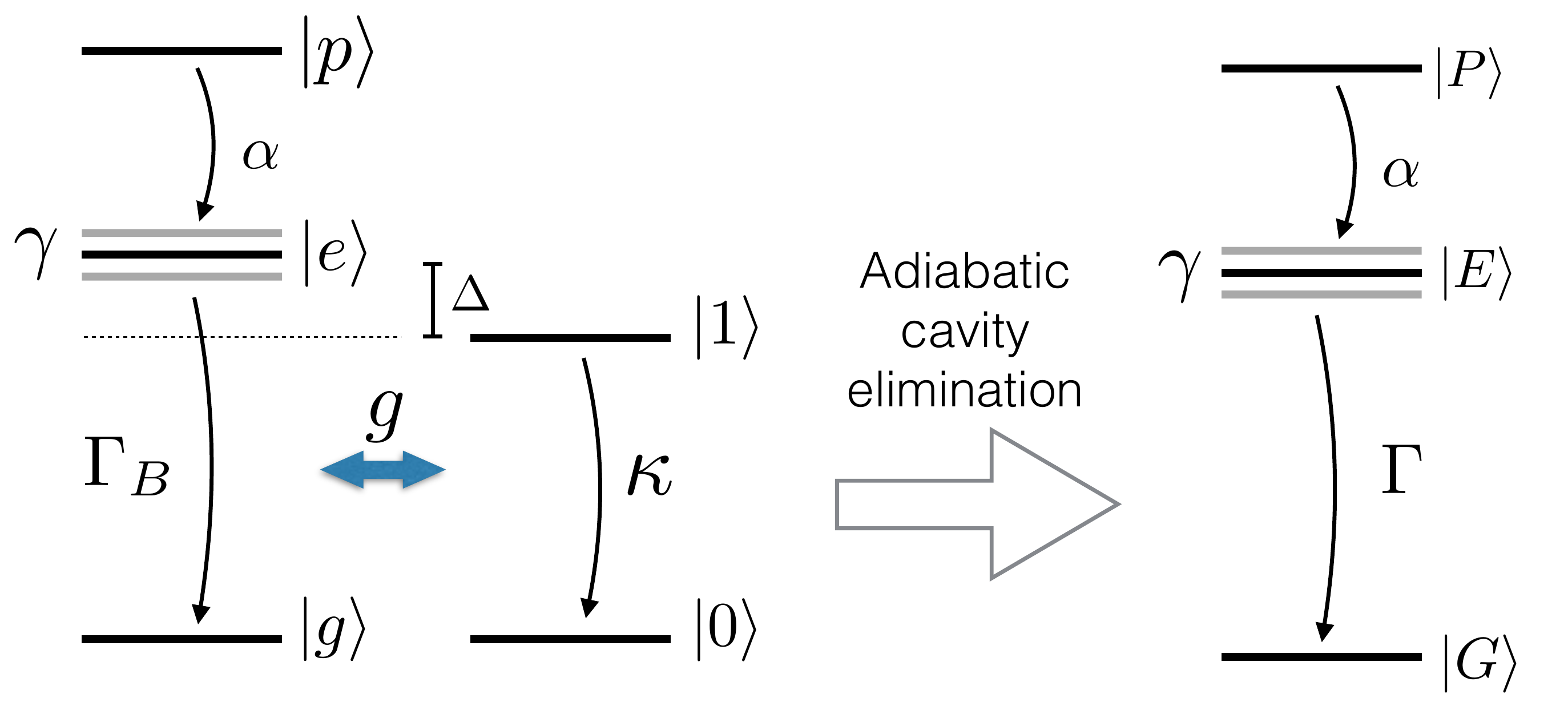}
\caption{Schematic diagram of the system under consideration; a quantum dot with pump-level $\ket{p}$, excited state $\ket{e}$, and ground state 
$\ket{g}$ couples to a cavity mode with strength $g$ and detuning $\Delta$. We consider only the zero photon and one photon manifolds of 
the cavity mode, which decays with rate $\kappa$. The $\ket{p}\to\ket{e}$ transition has rate $\alpha$, while the spontaneous emission 
process of $\ket{e}\to\ket{g}$ has background rate $\Gamma_B$. The excited state undergoes pure-dephasing with rate $\gamma$.  
Adiabatic cavity elimination results in an effective three-level system with a modified spontaneous emission rate $\Gamma$.}
\label{system}
\end{center}
\end{figure}

We now develop a master equation which will allow us to derive an analytic expression for  
$\smash{g^{(2)}_{\mathrm{HOM}}(\tau_D)}$ in the presence of time-jitter and pure-dephasing. 
We follow Ref.~\cite{Kaer2013} and model the quantum dot (QD) as a three-level-system, 
with crystal ground state $\ket{g}$, single exciton state $\ket{e}$, and pump level 
$\ket{p}$, having energies $\hbar\w_g$, $\hbar\w_e$ and $\hbar\w_p$ respectively. The 
cavity mode is described by creation and annihilation operators $c^{\dagger}$ and 
$c$, and has frequency $\w_c$. The system is depicted in Fig.~({\ref{system}}). In a rotating frame the QD--cavity system is described 
by the Jaynes-Cummings Hamiltonian
\begin{align}
H_{\mathrm{JC}}=\hbar\Delta\ketbra{e}{e}+\hbar g (\ketbra{e}{g} c+\ketbra{g}{e}c^{\dagger}),
\end{align}
where $\Delta=(\w_e-\w_g)-\w_c$ is the detuning of the QD transition from the cavity mode, and 
$g$ is the QD--cavity coupling strength. Relaxation processes are added using the Lindblad 
formalism~\cite{b+p}, and the master equation describing the QD--cavity degrees of freedom $\rho$ becomes 
\begin{align}
\der{\rho}{t}=-&\sfrac{i}{\hbar}[H_{\mathrm{JC}},\rho]+\big(L_{\kappa}(c)+L_{\Gamma_B}(\ketbra{g}{e})
\nonumber\\
+&L_{2\gamma}(\ketbra{e}{e})+L_{\alpha}(\ketbra{e}{p})\big)\rho
\end{align}
where the Lindblad operators satisfy 
$L_{\gamma}(A)\rho=\gamma\big(A\rho A^{\dagger}-\frac{1}{2}\{A^{\dagger}A,\rho\}\big)$, with  
$\alpha$ and $\kappa$ the decay rates of the pump-level and cavity respectively. 
The background spontaneous 
emission rate of the QD is $\Gamma_B$, and the rate $\gamma$ describes pure-dephasing of the QD excited state level.

In the limit of weak QD--cavity coupling and/or strong cavity decay, the cavity can be adiabatically 
eliminated from equations of motion describing our system. Formally, we require 
$\gamma_{\mathrm{tot}}\gg \Delta,\Gamma,g$ with 
$\gamma_{\mathrm{tot}}=\gamma+\frac{1}{2}(\kappa+\Gamma)$, and provided we consider the initial 
state $\rho(0)=\ketbra{p}{p}\ketbra{0}{0}$ with $\ket{0}$ the vacuum state of the cavity mode, the 
dynamics can be well approximated by the master equation~\cite{Kaer2013}
\begin{align}
&\der{\rho}{t}=-\sfrac{i}{\hbar}[\Delta\ketbra{E}{E},\rho]+ \nonumber \\
&\big(L_{\Gamma}(\ketbra{G}{E})
+L_{2\gamma}(\ketbra{E}{E})+L_{\alpha}(\ketbra{E}{P})\big)\rho,
\label{MEFinal1}
\end{align}
which is Eq.~({\ref{MEFinal}}) in the main text. 

\subsection{Photon Indistinguishability for the QD--Cavity System}

We now use Eq.~({\ref{MEFinal1}}) 
to calculate the two-photon interference probability, Eq.~({\ref{g2HOM}}). 
To proceed, we note that in the far field 
we can make the replacement $a(t)\to \sigma(t)$ with $\sigma=\ketbra{G}{E}$~\cite{kiraz04} in Eq.~({\ref{g2HOM}}). 
Then, to calculate the second order correlation function 
$\smash{G^{(2)}_{\mathrm{HBT}}(t_1,t_2)}=\frac{1}{4}\langle a^{\dagger}(t_1) a^{\dagger}(t_2) a(t_1) a(t_2)\rangle
\to \frac{1}{4}\langle \sigma^{\dagger}(t_1) \sigma^{\dagger}(t_2) \sigma(t_1) \sigma(t_2)\rangle$, 
we make use of quantum regression theorem to write~\cite{carmichael}
\beq
\pder{}{\tau}G^{(2)}_{\mathrm{HBT}}(t,t+\tau)=-\Gamma\, G^{(2)}_{\mathrm{HBT}}(t,t+\tau).
\eeq
For $\tau=0$ we find $G^{(2)}_{\mathrm{HBT}}(t,t)=0$ since $\sigma^2=(\sigma^{\dagger})^2=0$, and as such 
$G^{(2)}_{\mathrm{HBT}}(t,t+\tau)=0$ and we can set $g^{(2)}(0)=0$ in Eq.~({\ref{g2HOM}}). This reflects 
that for the theory presented here we have strictly one (or less) excitation in the system at any time. 

We now calculate the two-photon coalescence probability expressed in Eq.~({\ref{Ctau}}). To 
begin we consider the uncorrelated probability  
$\mathcal{G}^{(2)}(t_1,t_2)\to \av{\sigma^{\dagger}(t_1)\sigma(t_1)}\av{\sigma^{\dagger}(t_2)\sigma(t_2)}$. 
The quantity $\av{\sigma^{\dagger}(t)\sigma(t)}=\mathrm{Tr}(\rho(t)\sigma^{\dagger}\sigma)$ is just the excited 
state population at time $t$, and from Eq.~({\ref{MEFinal1}}) we have 
\beq
\av{\sigma^{\dagger}(t)\sigma(t)}=\Theta(t)\frac{\alpha}{\Gamma-\alpha}\left(\e^{-\alpha t}-\e^{-\Gamma t}\right)
\eeq
where the Heaviside theta function ($\Theta(t)=0$ for $t<0$ and $\Theta(t)=1$ for $t>0$) has been introduced to ensure 
no excitations are present before emission events. From the quantum regression theorem 
the first order correlation function $G^{(1)}(t,t+\tau)$ obeys the equation of motion
\beq
\pder{}{\tau}G^{(1)}(t,t+\tau)=-\Big(\gamma+\sfrac{1}{2}\Gamma+i \Delta\Big) G^{(1)}(t,t+\tau)
\eeq
with initial condition $G^{(1)}(t,t)=\av{\sigma^{\dagger}(t)\sigma(t)}$, which gives 
\begin{align}
G^{(1)}(t,t+\tau)=
\av{\sigma^{\dagger}(t)\sigma(t)}\e^{-(\gamma+\frac{1}{2}\Gamma+i \Delta)|\tau|}.
\end{align}
Finally, performing the integrals in Eq.~({\ref{g2HOM}}) we arrive at Eq.~({\ref{HOMDipalpha}}) in the main text.

\subsection{Exciton--phonon coupling}
\label{EPC}

To explore the influence of phonons seen in our data, a weak exciton--phonon coupling 
time convolutionless master equation technique is used~\cite{Kaer2013}. To second 
order in the exciton--phonon coupling strength, and within the Born-Markov approximation, the master equation 
for the complete QD--cavity system (i.e. before adiabatic elimination) becomes
\begin{align}
\der{\rho}{t}&=-\sfrac{i}{\hbar}[H_{\mathrm{JC}},\rho]+\big(L_{\kappa}(c)+L_{\Gamma}(\ketbra{g}{e})\nonumber\\
&+L_{2\gamma}(\ketbra{e}{e})+L_{\alpha}(\ketbra{e}{p})\big)\rho+\mathcal{K}_{\mathrm{ph}}(\rho),
\label{MEPhonons}
\end{align}
where the new phonon-induced dissipator is given by 
\begin{align}
\mathcal{K}_{\mathrm{ph}}(\rho)=
-\!\!\int_0^{\infty}\!\!\! d s \mathrm{Tr}_{\mathrm{ph}}\big[ H_I , [\tilde{H}_I (-s), \rho(t) \rho_{\mathrm{ph}}]\big],
\label{Kph}
\end{align}
where $\mathrm{Tr}_{\mathrm{ph}}$ denotes a trace over the phonon modes. The interaction Hamiltonian is written
\beq
H_I=\ketbra{e}{e}\sum_k g_k(b_k^{\dagger}+b_k)
\eeq
where $b_k^{\dagger}$ is a creation operator for a phonon mode with wave-vector $k$, and $g_k$ describes its coupling 
strength to the QD exciton. The interaction picture interaction Hamiltonian is defined by 
$\tilde{H}(-s)=\e^{-i H_0 s}H_I\e^{i H_0 s}$, where $H_0=H_{\mathrm{JC}}+H_{\mathrm{ph}}$, with 
phonon Hamiltonian $\smash{H_{\mathrm{ph}}=\sum_k\w_k b_k^{\dagger}b_k}$ and $\w_k$ the frequency 
of mode $k$. Finally, we assume a thermal state for the phonon density operator: 
$\rho_{\mathrm{ph}}=\e^{-\beta H_{\mathrm{ph}}}/\mathrm{Tr}_{\mathrm{ph}}(\e^{-\beta H_{\mathrm{ph}}})$, 
with $\beta=\hbar/k_B T$ and $T$ the sample temperature.

The strength of the QD--phonon coupling 
is characterised by the spectral density, defined as $J(\w)=\sum_k g_k^2 \delta(\w-\w_k)$, and 
which for excitons in QDs has been shown to be adequately described by the function 
\beq
J(\w)= \eta\, \w^3 \exp\big[-(\w/\w_c)^2\big],
\eeq
where $\eta$ captures the overall strength of the interaction determined by material parameters, 
and $\w_c$ is the photon cut-off frequency~\cite{ramsay10}. 
The behaviour of the phonon dissipator in Eq.~({\ref{Kph}}) in different parameter regimes has been discussed in 
detail elsewhere~\cite{Kaer2013,Kaer13_2,Kaer2014}. The parameters used to obtain 
improved fits to the data in the main text (the solid orange curves in Fig. \ref{HOMFits}) are  
$\eta=0.032~\mathrm{meV}^{-2}$ and $\w_c=1.3~\mathrm{meV}$, 
while the constant pure-dephasing rate was reduced to $\gamma=1~\mu\mathrm{eV}$. These parameters correspond 
to phonons contributing approximately $40\%$ of the dephasing on-resonance. We note 
that the other parameters in the model were adjusted to $1/\Gamma_B=730~\mathrm{ps}$ and $g=34~\mu\mathrm{eV}$
in order that the $T_1$ times as a function of detuning were well reproduced.

The density operator $\rho$ entering Eq.~({\ref{MEPhonons}}) contains both 
QD and cavity degrees of freedom. When relating the field operator $a(t)$ to the QD--cavity system, 
we have a choice to consider QD emission or cavity emission, making respectively the replacements 
$a\to c$ or $a\to\ketbra{g}{e}$ in the field correlation functions. 
Our data was better described by cavity emission, which we attribute 
to the high Purcell factor of our QD--cavity system.


\end{document}